\documentclass[prb,preprint,showpacs]{revtex4}

\usepackage{amsmath,amssymb}

\usepackage{graphicx}

\begin{document}

\newtheorem{Lemma}{Lemma}
\newtheorem{Theorem}{Theorem}

\hyphenation{super-lat-tice semi-con-ductor}

\title{Nonlinear electron and spin transport in semiconductor superlattices }

\author{L. L. Bonilla}
\affiliation{G. Mill\'an Institute of Fluid Dynamics, Nanoscience and
Industrial Mathematics, Universidad Carlos III de Madrid, Avenida de la Universidad 30, 
28911 Legan{\'e}s, Spain ({\tt bonilla@ing.uc3m.es}).}
\author{L. Barletti }
\affiliation{Dipartimento di Matematica ``Ulise Dini''; Universit\`a di Firenze; 
Viale Morgagni 67/A; 50134 Firenze, Italy ({\tt barletti@math.unifi.it}).} 
\author{M. Alvaro} 
\affiliation{G. Mill\'an Institute of Fluid Dynamics, Nanoscience and
Industrial Mathematics, Universidad Carlos III de Madrid, Avenida de la Universidad 30, 
28911 Legan{\'e}s, Spain ({\tt mariano.alvaro@uc3m.es})} 

\date{\today}

\setcounter{page}{1}
\renewcommand{\thefootnote}{\arabic{footnote}}
\renewcommand{\theequation}{\arabic{section}.\arabic{equation}}
\newcommand{\fin}{\newline \rule{2mm}{2mm}}
\def\RR{\mathbb{R}}
\def\pRR{\mathbb{R}}
\def\ZZ{\mathbb{Z}}

\begin{abstract}
Nonlinear charge transport in strongly coupled semiconductor superlattices is described by
Wigner-Poisson kinetic equations involving one or two minibands. Electron-electron 
collisions are treated within the Hartree approximation whereas other inelastic collisions are 
described by a modified BGK (Bhatnaghar-Gross-Krook) model. The hyperbolic limit is such that the
collision frequencies are of the same order as the Bloch frequencies due to the electric
field and the corresponding terms in the kinetic equation are dominant. In this limit, 
spatially nonlocal drift-diffusion balance equations for the miniband populations and the 
electric field are derived by means of the Chapman-Enskog perturbation technique. 
For a lateral superlattice with spin-orbit
interaction, electrons with spin up or down have different energies and their corresponding
drift-diffusion equations can be used to calculate spin-polarized currents and electron spin
polarization. Numerical solutions show stable self-sustained oscillations of the current and 
the spin polarization through a voltage biased lateral superlattice thereby providing an
example of superlattice spin oscillator. 
\end{abstract}

\maketitle

\pagestyle{myheadings}
\thispagestyle{plain}
\markboth{L.~L.~BONILLA ET AL}{
NONLINEAR ELECTRON AND SPIN TRANSPORT IN SUPERLATTICES}

\setcounter{equation}{0}
\section{Introduction}\label{sec:intro}
\begin{figure}
\begin{center}
\includegraphics[width=11.cm,angle=0]{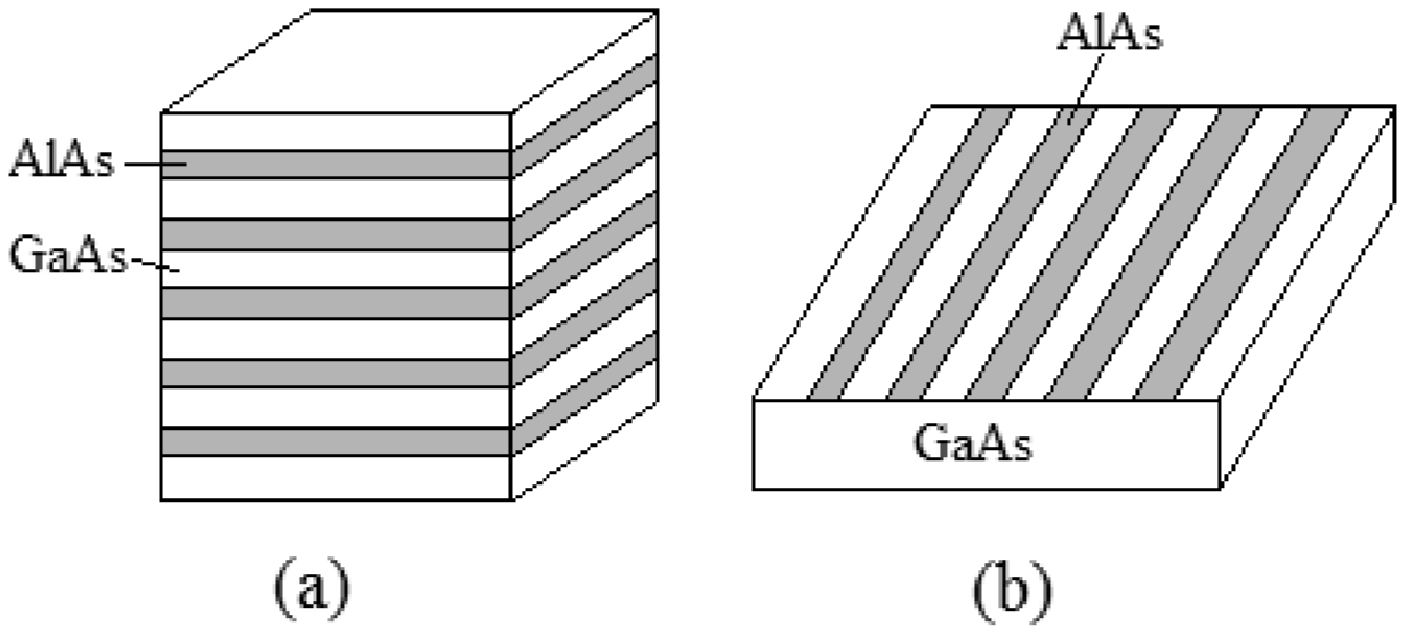}
\vspace{0.2cm}
\caption{(a) Schematic drawing of a superlattice. (b) A lateral superlattice.}
\label{fig1}
\end{center}
\end{figure}

Semiconductor superlattices are essential ingredients in fast nanoscale oscillators, quantum
cascade lasers and infrared detectors. Quantum cascade lasers are used to monitor 
environmental pollution in gas emissions, to analyze breath in hospitals and in many other
industrial applications \cite{BGr05}. A superlattice (SL) is a convenient approximation to a 
quasi-one-dimensional crystal that was originally proposed by Esaki and Tsu to observe 
Bloch oscillations, i.e., the periodic coherent motion of electrons in a miniband in the 
presence of an applied electric field. Fig.~\ref{fig1}(a) shows a simple realization of a 
$N$-period SL. Each period of length $l$ consists of two layers of semiconductors with 
different energy gaps but with similar lattice constants. The SL lengths in the lateral 
directions, $L_{y}$ and $L_{z}$ are much larger than $l$, typically tens of microns compared
to about ten nanometers. The energy profile of the conduction band of this SL can be
modeled as a succession of square quantum wells and barriers along the $x$ direction 
(Kronig-Penney model) and, for a n-doped SL, we do not have to consider the valence band. 
A different quasi-1D crystal called a lateral superlattice (LSL) is shown in 
Fig.~\ref{fig1}(b). In this case, a periodic structure is formed on the top surface of a 
quantum well (QW), so that $L_{z}$ is of the order of $l$ and $L_{y}\gg l$. The wave 
functions of a single electron in the conduction band of a SL can be expanded in terms of 1D 
Bloch wave functions times plane waves
\begin{eqnarray}
{1\over\sqrt{S}}\, e^{ik_{y}y}\psi(z)\, e^{ikx} u_{\nu}(x,k), \label{1.1}\\
\psi(z)= \left\{\begin{array}{cc}
 e^{ik_{z}z}, &\mbox{for a SL,}\\
 \psi_{n}(z), &\mbox{for a LSL,}\end{array}\right. \label{1.2}
\end{eqnarray}
where $\nu$ is the miniband index and $n$ is the energy level of the quantum well in 
the case of a LSL. The function $u_{\nu}(x,k)$ is $l$-periodic in $x$ and 
$2\pi/l$-periodic in $k$. $S$ is the area of the lateral cross section, equal to $L_{y}L_{z}$
for a rectangular cross section. 

Many interesting nonlinear phenomena have been observed in voltage biased SL comprising
finitely many periods, including self-oscillations of the current through the SL due to motion 
of electric field pulses, multistability of stationary charge and field profiles, and so on 
\cite{BGr05}. It is important to distinguish between strongly and weakly coupled SLs 
depending on the coupling between their component QWs. Roughly speaking, if barriers are
narrow, QWs are strongly coupled and we can use the electronic states (\ref{1.1})
as a convenient basis in a quantum kinetic description. The resulting reduced balance 
equations for electron density and electric field are partial differential equations (which may 
be nonlocal, as we shall see in this paper). On the other hand, for SLs having wide barriers, 
their QWs are weakly coupled and the electronic states of a single well provide a good basis 
in a quantum kinetic description, replacing the Bloch functions $e^{ikx}u_{\nu}(x,k)$
in (\ref{1.1}). In this case, the balance equations are spatially discrete and phenomena such 
as multistability of stationary field profiles, formation and pinning of electric field domains, 
etc are theoretically predicted and observed in experiments. See the review \cite{BGr05}. 
Another promising field of applications is spintronics. 
Electrons in SLs having at least one period doped with magnetic impurities and subject to a 
static magnetic field can be distinguished by their spin because the magnetic field splits each 
miniband in two having different spin-dependent energy \cite{SMP02}. Recently a SL of 
this type has been proposed as a spin oscillator producing spin polarized oscillatory currents 
and able to inject polarized electrons in a contact \cite{Bon07}. Alternatively, materials 
displaying strong spin-orbit effects can be used as spintronic devices without having to apply 
magnetic fields; cf.\ the case of the LSL considered in Ref.~\cite{KBB05}. In this paper,
we will show that a LSL can be used as a spin oscillator.

This paper presents systematic derivations of quantum balance equations for SLs with two 
populated minibands, and it shows that their numerical solutions may predict space and
time-dependent nonlinear phenomena occurring in these materials. Our methods can be used 
in 3D crystals, but their application to 1D structures such as SLs and LSLs leads to simpler 
equations that are less costly to solve. Although nonlinear charge transport in SLs has been 
widely studied in the last decade (see the reviews \cite{BGr05,Pla04,Wac02}), systematic 
derivations of tractable balance equations for miniband populations and electric field are 
scarce. One reason is that quantum kinetic equations are nonlocal in space and their collision 
terms may be nonlocal in space and time \cite{jauho,Wac02}. Using them to analyze space 
and time-dependent phenomena such as wave propagation or self-sustained oscillations
is problematic. In fact, only extremely simple solutions of general quantum kinetic equations 
(such as thermal equilibrium, disturbances thereof due to weak external fields and so on) are 
known, theoretical analysis of these equations is lacking and numerical solutions describing 
spatio-temporal phenomena are not available. One way to proceed is to adopt 
simple collision models similar to the Bhatnagar-Gross-Krook (BGK) collision model for 
classical kinetic theory \cite{BGK54}. We discuss in this paper how to implement a BGK 
collision model for a quantum kinetic equation that is simple to handle yet keeps an important 
quantum feature such as the broadening of energy levels \cite{BEs05}. Once we have a 
quantum kinetic equation for a sufficiently general SL having two minibands, we implement 
a Chapman-Enskog perturbation procedure to derive the sought balance equations and solve 
them numerically for realistic SL configurations.

Previous to this work, Lei and coworkers derived quantum hydrodynamic equations describing
SL having only one miniband \cite{LTi84,Lei95}. They use a closure assumption to close 
a hierarchy of moment equations. For the case of quantum particles in an arbitrary external 
three-dimensional potential, Degond and Ringhofer \cite{DRi03} have used a similar 
procedure to derive balance equations. They close the system of moment equations by means 
of a local equilibrium density obtained by maximizing entropy. The Chapman-Enskog 
method has been used to derive drift-diffusion equations for single-miniband SLs described 
by semiclassical \cite{BEP03} and quantum kinetic equations \cite{BEs05}. Earlier, 
Cercignani, Gamba and Levermore used the Chapman-Enskog method to derive balance 
equations for a semiclassical BGK-Poisson kinetic description of a semiconductor with one
parabolic band under strong external bias \cite{CGL01}. 

The rest of this paper is as follows. In Section \ref{sec:2}, we review the simpler
case of nonlinear electron transport in a  strongly coupled n-doped SL having only one 
populated miniband \cite{BEs05}. Starting with a kinetic equation for the Wigner function, 
we use the Chapman-Enskog perturbation method to derive balance equations for the electron 
density and the electric field. When these equations are solved numerically for a dc voltage
biased SL with finitely many QWs and realistic parameter values, stable self-sustained 
oscillations of the current through the SL are found among their solutions, in agreement 
with experimental observations \cite{BEs05}. Sections \ref{sec:3} to \ref{sec:5}
contain the main results of the present work. In Section \ref{sec:3}, we describe a
SL having two populated minibands by proposing a kinetic equation for the Wigner matrix.
In Section \ref{sec:4}, we derive balance equations for the miniband electron populations 
and the electric field, using an appropriate Chapman-Enskog method and a tight-binding 
approximation to obtain explicit formulas. 
The case of a LSL having strong Rashba spin-orbit interaction \cite{rashba} is important for 
spintronic applications and has been considered in Section \ref{sec:5}. We derive and
solve numerically the resulting balance equations. Novel self-sustained oscillations of the
spin current and polarization are obtained for appropriate values of the parameters. Finally
Section \ref{sec:6} contains our conclusions and some technical matters are relegated to the 
Appendix. 

\setcounter{equation}{0}
\section{Single miniband superlattice}
\label{sec:2}
The Wigner-Poisson-Bhatnagar-Gross-Krook (WPBGK) system for 1D electron transport in 
the lowest miniband of a strongly coupled SL is:
\begin{eqnarray} \label{1}
&&{\partial f\over \partial t} + \frac{i}{\hbar}\left[ {\cal E}\left(k + 
{1\over 2i}{\partial\over \partial x}\right) - {\cal E}\left(k - {1\over 2i}
{\partial\over \partial x}\right)\right]\, f \\
&&\quad +  {ie\over \hbar} \,\left[W
\left(x+{1\over 2i}{\partial\over \partial k},t\right) - W\left(x-{1\over 
2i}{\partial\over \partial k},t\right) \right]\, f \nonumber\\
&&\quad = Q[f]\equiv - \nu_{en}\,  \left(f - f^{FD}
\right) - \nu_{\rm imp}\, {f(x,k,t) - f(x,-k,t)\over 2}  ,  \nonumber\\
&&\varepsilon\, {\partial^2 W\over\partial x^2} = {e\over l}\, 
(n-N_{D}),  \label{2}\\   
&& n(x,t) = { l\over 2\pi} \int_{-\pi/l}^{\pi/l} f(x,k,t) dk =
{ l\over 2\pi} \int_{-\pi/l}^{\pi/l} f^{FD}(k;n(x,t)) dk,\quad \label{3}\\
&& f^{FD}(k;n) = {m^{*}k_{B}T\over\pi\hbar^2}\,\int_{-\infty}^\infty
\ln\left[1+ \exp\left({\mu - E\over k_{B}T}\right)\right]\, {\sqrt{2}\,\Gamma^3/
\pi\over [E-{\cal E}_{1}(k)]^4+\Gamma^4}\, dE. \label{4}
\end{eqnarray} 
Here $f$, $n$, $N_{D}$, ${\cal E}(k)$, $d_{B}$, $d_{W}$, $l=d_{B}+d_{W}$, $W$, 
$\varepsilon$, $m^*$, $k_{B}$, $T$, $\Gamma$, $\nu_{en}$, $\nu_{\rm imp}$ and $-e<0$ 
are the one-particle Wigner function, the 2D electron density, the 2D doping density, the 
miniband dispersion relation, the barrier width, the well width, the SL period, the electric 
potential, the SL permittivity, the effective mass of the electron in the lateral directions, the 
Boltzmann constant, the lattice temperature, the energy broadening of the equilibrium 
distribution due to collisions \cite{KBa62} (page 28 ss), the frequency of the inelastic 
collisions responsible for energy relaxation, the frequency of the elastic impurity collisions 
and the electron charge, respectively. 

The left-hand side of Eq.\ (\ref{1}) can be straightforwardly derived from the 
Schr\"od\-inger-Poisson equation for the wave function in the miniband using the definition 
of the 1D Wigner function \cite{BEs05}:
\begin{eqnarray}
f(x,k,t)= {2l\over S}\sum_{j=-\infty}^\infty \int_{\mathbb{R}^2}\langle
\psi^\dagger(x+jl/2,y,z,t)\psi (x-jl/2,y,z,t)\rangle e^{ijkl} d\mathbf{x}_{\perp}   
\label{wigner}
\end{eqnarray}
 (the second quantized wave function $\psi(x,\mathbf{x}_{\perp},t) =\sum_{q,\mathbf{q
 }_{\perp}} a(q,q_{\perp},t) \phi_{q}(x) e^{i\mathbf{q}_{\perp}\cdot\mathbf{x
 }_{\perp}}$, $\mathbf{x}_{\perp}=(y,z)$, is a superposition of the Bloch states 
 corresponding to the miniband and $S$ is the SL cross section\,Ê\cite{BEs05}). The right 
 hand side in Eq.~(\ref{1}) is the sum of $-\nu_{e}\left(f - f^{FD} \right)$, which 
 represents energy relaxation towards a 1D effective Fermi-Dirac (FD) distribution $f^{FD}
 (k;n)$ (local equilibrium), and $-\nu_{i} [f(x,k,t) -f(x,-k,t)]/2$, which accounts for impurity 
 elastic collisions \cite{ BEP03}. For simplicity, the collision frequencies $\nu_{e}$ and 
$\nu_{i}$ are fixed constants. Exact and FD distribution functions have the same electron 
density, thereby preserving charge continuity as in the classical BGK collision 
models~\cite{BGK54}. The chemical potential $\mu$ is a function of $n$ resulting from 
solving equation (\ref{3}) with the integral of the collision-broadened 3D Fermi-Dirac 
distribution over the lateral components of the wave vector $(k,\mathbf{k}_{\perp})=(k,
k_{y},k_{z})$:
\begin{eqnarray}
&& f^{FD}(k;n) = \int_{-\infty}^\infty {D_{\Gamma}\left(E-{\cal E}_{1}(k)
\right)\over 1+\exp\left({E-\mu\over k_{B}T}\right)}\,  dE, \label{4.1}\\
&& D_{\Gamma}(E) = {2\over(2\pi)^2} \int_{\mathbb{R}^2} \delta_{\Gamma}
\left({\hbar^2\mathbf{k}_{\perp}^2\over 2 m^*}-E\right)\, d\mathbf{k}_{\perp}
={m^*\over\pi\hbar^2} \int_0^{\infty} \delta_{\Gamma}(E_{\perp}-E)\, 
dE_{\perp}.  \label{4.2}
\end{eqnarray}
Using the residue theorem for a line-width:
\begin{eqnarray}
\delta_{\Gamma}(E) &=& {\sqrt{2}\,\Gamma^3/\pi\over\Gamma^4+E^4}, \label{4.3}
\end{eqnarray} 
(\ref{4.2}) yields
\begin{eqnarray}
D_{\Gamma}(E)&=& {m^*\over\pi
\hbar^2}\left\{1 +{1\over 4\pi}\,\ln\left[{E^2+\sqrt{2}\Gamma E+\Gamma^2
\over E^2-\sqrt{2}\Gamma E+\Gamma^2}\right] \right.\label{4.4}\\
&-& {\theta(\sqrt{2}|E|-\Gamma)\over 2\pi}
\left[2\pi -\arctan\left({\Gamma\over\sqrt{2}|E|+\Gamma}\right) - 
\arctan\left({\Gamma\over\sqrt{2}|E|-\Gamma}\right)\right]\nonumber\\
&-& {\theta(\Gamma-\sqrt{2}|E|)\over 2\pi}\left[
\pi+\arctan\left({\Gamma\over\sqrt{2}E+\Gamma}\right) - \arctan\left({\Gamma
\over \Gamma-\sqrt{2}E}\right)\right] \nonumber\\
&-&\left. {\theta(\sqrt{2}E-\Gamma)\over 2\pi}
\left[\arctan\left({\Gamma\over\sqrt{2}E+\Gamma}\right) + 
\arctan\left({\Gamma\over\sqrt{2}E-\Gamma}\right)\right] \right\}, \nonumber
\end{eqnarray}
which is equivalent to Eq.\ (\ref{4})\footnote{Integrate (\ref{4.1}) by parts using
(\ref{4.4}).}. Here $\theta(E)$ is the Heaviside unit step function. As $\Gamma\to 0+$, 
the line-width (\ref{4.3}) tends to the delta function $\delta(E)$, $D_{\Gamma}(E)$ tends 
to the 2D density of states, $D(E)= m^*\theta(E)/(\pi\hbar^2)$, and $f^{FD}$ tends to the 
3D Fermi-Dirac distribution function integrated over the lateral wave vector $\mathbf{k}_{
\perp}$. In Ref.~\cite{BEs05}, a Lorentzian line-width was used instead of (\ref{4.3}) 
and the integral over $E$ in (\ref{4.1}) extended from 0 to $\infty$. The integral with the 
Lorentzian function is not convergent in $E=-\infty$, which is why we prefer using 
convolution with the ``super-Lorentzian'' function (\ref{4.3}) in this work. The integration 
in (\ref{4.2}) cannot be carried out explicitly for other standard line-widths such as a 
Gaussian or a hyperbolic secant. This unnecessarily complicates the numerical integration of 
the balance equations we will obtain later. Note that, following Ignatov and Shashkin 
\cite{ISh87}, we have not included the effects of the electric potential 
in our Fermi-Dirac distribution. These model equations can be improved by including 
scattering processes with change of lateral momentum and an electric field-dependent local 
equilibrium. However the resulting model could only be treated numerically and the 
qualitative features of our derivation and of the nonlocal drift-diffusion equation would be 
lost in longer formulas.

A different way to introduce a quantum BGK collision model is to define a local 
equilibrium density matrix operator by minimizing quantum entropy (defined with the 
opposite sign of the convention that is usual in physics) under constraints giving
the electron density and energy density in terms of the density matrix. The resulting
expression involves an inverse Wigner transform and another transform is needed to deduce
the local equilibrium Wigner function $f^{FD}$ entering the BGK formula \cite{DRi03}. 
This $f^{FD}$ is nonlocal in space and can only be found by solving some partial differential 
equation \cite{DRi03}. As a model for quantum collisions \cite{jauho,Wac02}, the 
resulting quantum BGK model is not realistic, in the same way as the original BGK model is 
not a realistic model for classical collisions. Moreover, the implicit manner in which the 
model is defined defeats the main asset of the classical BGK collision model: its simplicity, 
that makes it possible to obtain results analytically. Thus we prefer to introduce a BGK model 
that can be handled more easily and still incorporates quantum effects. The most important 
quantum effect affecting the collision term is the broadening of energy levels due to 
scattering, $\Gamma\approx\hbar/\tau$ (where $\tau$ is the lifetime of the level)
\cite{KBa62}, and this is taken phenomenologically into account by the convolution with 
the line-width function (\ref{4.3}) in (\ref{4.1}). In the semiclassical limit ``$\hbar\to 
0$'', $\Gamma\to 0$ and we recover the semiclassical FD distribution.

The WPBGK system (\ref{1}) to (\ref{4}) should be solved for a Wigner function which 
is $2\pi/l$-periodic in $k$ and satisfies appropriate initial and boundary conditions. It is 
convenient to derive the charge continuity equation and a nonlocal Amp\`ere's law for the 
current density. The Wigner function $f$ is periodic in $k$; its Fourier expansion is 
\begin{equation}
f(x,k,t)= \sum_{j=-\infty}^{\infty} f_{j}(x,t)\, e^{ijkl}.\label{5}
\end{equation}
Defining $F=\partial W/\partial x$ ({\em minus} the electric field) and the average
\begin{equation}
\langle F\rangle_{j}(x,t) = {1\over jl} \int_{-jl/2}^{jl/2} F(x+s,t)\, ds, \label{6}
\end{equation}
it is possible to obtain the following equivalent form of the Wigner equation \cite{BEs05}
\begin{eqnarray}
{\partial f\over \partial t} + \sum_{j= -\infty}^{\infty} {i jl\over\hbar}\,
e^{ijkl} \left( {\cal E}_{j}\, {\partial\over \partial x}\langle f\rangle_{j}
+ e\, \langle F\rangle_{j}\, f_{j} \right) = Q[f].
\label{7}
\end{eqnarray}
Here the nonzero Fourier coefficients of the dispersion relation are simply ${\cal E}_{0}=
\Delta/2$ and ${\cal E}_{\pm 1}= -\Delta/4$ for the tight-binding dispersion relation 
${\cal E}(k)=\Delta\, (1- \cos kl)/2$ ($\Delta$ is the miniband width), which yields a 
miniband group velocity $v(k)= {\Delta l\over 2\hbar}\sin kl$.
Integrating this equation over $k$ yields the charge continuity equation
\begin{eqnarray}
{\partial n\over \partial t} + {\partial\over \partial x}\, \sum_{j= 1}^{\infty
}{2jl\over\hbar} \left\langle \mbox{Im}({\cal E}_{-j} f_{j})\right\rangle_{j}=0.  
\label{8}
\end{eqnarray}
Here we can eliminate the electron density by using the Poisson equation
and then integrate over $x$, thereby obtaining the nonlocal Amp\`ere's law
for the total current density $J(t)$:
\begin{eqnarray}
\varepsilon\, {\partial F\over \partial t} + {2e\over\hbar}\,
\sum_{j= 1}^{\infty} j\langle \mbox{Im} ({\cal E}_{-j} f_{j})\rangle_{j}
= J(t).   \label{9}
\end{eqnarray}

To derive the QDDE, we shall assume that the electric field contribution in Eq.\ (\ref{7})
is comparable to the collision terms and that they dominate the other terms ({\em the
hyperbolic limit}) \cite{BEP03}. Let $v_{M}$ and $F_{M}$ be the electron velocity
and field positive values at which the (zeroth order) drift velocity reaches its maximum.
In this limit, the time $t_{0}$ it takes an electron with speed $v_{M}$
to traverse a distance $x_{0}=\varepsilon F_{M}l/(eN_{D})$, over
which the field variation is of order $F_{M}$, is much longer than the mean free time
between collisions, $\nu_{e}^{-1}\sim \hbar/(eF_{M}l)=t_{1}$. We therefore define the
{\em small parameter} $\lambda=t_{1}/t_{0}=\hbar v_{M}N_{D}/(\varepsilon 
F_{M}^2 l^2)$ and formally multiply the first two terms on the left side of (\ref{1}) or
(\ref{7}) by $\lambda$~\cite{BEP03,BEs05}. The result is 
\begin{eqnarray}
\lambda\left({\partial f\over \partial t} + \sum_{j= -\infty}^{\infty} {i jl\over
\hbar}\, e^{ijkl} {\cal E}_{j}\, {\partial\over \partial x}\langle f\rangle_{j}
\right) = Q[f] - \sum_{j= -\infty}^{\infty} {i ejl\over\hbar}\, e^{ijkl}\langle F
\rangle_{j}\, f_{j}.   \label{10}
\end{eqnarray}
The solution of Eq.\ (\ref{10}) for $\lambda=0$ is calculated in terms of its Fourier 
coefficients as
\begin{eqnarray}
f^{(0)}(k;F) = \sum_{j=-\infty}^{\infty} {(1-ij {\cal F}_{j}/\tau_{e})\,f^{FD}_{j}
\over 1 + j^2 {\cal F}_{j}^{2}}\, e^{ijkl}, \label{11}
\end{eqnarray}
where ${\cal F}_{j} = \langle F\rangle_{j}/F_{M}$, $F_M = {\hbar \over el} 
\sqrt{\nu_e (\nu_e+\nu_i)}$ and $\tau_e= \sqrt{(\nu_e + \nu_i) / \nu_e}$.

The Chapman-Enskog ansatz for the Wigner function is \cite{BEs05}:
\begin{eqnarray}
&& f(x,k,t;\lambda) = f^{(0)}(k;F) + \sum_{m=1}^{\infty} f^{(m)}(k;F)\, 
\lambda^{m} ,    \label{12}\\
&&  \varepsilon {\partial F\over\partial t} + \sum_{m=0}^{\infty}  J^{(m)}(F)\,
 \lambda^{m} = J(t).  \label{13}
\end{eqnarray}
The coefficients $f^{(m)}(k;F)$ depend on the `slow variables' $x$ and
$t$ only through their dependence on the electric field and the electron density. 
The electric field obeys a reduced evolution equation (\ref{13}) in which the functionals 
$J^{(m)}(F)$ are chosen so that the $f^{(m)}(k;F)$ are bounded and $2\pi/l$-periodic in 
$k$. After we keep the desired number of terms and set $\lambda=1$, Eq.\ (\ref{13})
is the QDDE provided by our perturbation procedure. 

Differentiating the Amp\`ere's law (\ref{13}) with respect to $x$, we obtain the charge 
continuity equation. Moreover the compatibility condition
\begin{eqnarray} 
\int_{-\pi/l}^{\pi/l} f^{(m)}(k;n) \, dk = {2\pi\over l}\, f^{(m)}_{0} = 0,  \quad
 m\geq 1, \label{14}  
\end{eqnarray}
is obtained by inserting the expansion (\ref{12}) into (\ref{3}). Inserting (\ref{12}) 
and (\ref{13}) in (\ref{10}), we find the hierarchy:
\begin{eqnarray} 
{\cal L} f^{(1)} &=&  \left.  -
{\partial  f^{(0)}\over \partial t}\right|_{0} + \sum_{j= -\infty}^{\infty} {i jl
 {\cal E}_{j}e^{ijkl}\over \hbar} {\partial\over \partial x}\langle f^{(0)}
 \rangle_{j}  \label{15}\\
 {\cal L} f^{(2)} &=&  \left.  - {\partial  f^{(1)}\over \partial t}\right|_{0}
 + \sum_{j= -\infty}^{\infty} {i jl {\cal E}_{j}e^{ijkl}\over \hbar} {\partial
 \over \partial x}\langle f^{(1)}\rangle_{j} \;
 - \left. {\partial \over \partial t} f^{(0)}\right|_{1}, \quad \label{16}
\end{eqnarray}
and so on. Here 
\begin{eqnarray} 
{\cal L} u(k) \equiv {ie\over\hbar}\sum_{-\infty}^{\infty} jl \langle F
\rangle_{j} u_{j} e^{ijkl} + \left( \nu_{e} + {\nu_{i}\over 2}\right) u(k) - {\nu_{i}
\over 2}\, u(-k), \label{17}
\end{eqnarray} 
and the subscripts 0 and 1 in the right hand side of these equations mean that $\varepsilon\,
\partial F/\partial t$  is replaced by $J - J^{(0)}(F)$ and by $-J^{(1)}(F)$, respectively.

The condition (\ref{14}) implies that 
\begin{eqnarray}
\int_{-\pi/l}^{\pi/l} {\cal L} f^{(m)} dk =0,\label{e18}
\end{eqnarray} 
for $m\geq 1$. Using this, the solvability conditions for the linear hierarchy of equations yield
\begin{eqnarray}
J^{(m)} = {2e\over\hbar}\,
\sum_{j= 1}^{\infty} j\langle \mbox{Im} ({\cal E}_{-j} f^{(m)}_{j})\rangle_{j},
\label{19}
\end{eqnarray}
which can also be obtained by insertion of Eq.\ (\ref{12}) in (\ref{9}).

Particularized to the case of the tight-binding dispersion relation and $\Gamma=0$ in the
Fermi-Dirac distribution (\ref{4}), the leading order of the Amp\`ere's law (\ref{13}) is
\begin{eqnarray}
\varepsilon {\partial F\over\partial t} + {e v_{M}\over l}\, \langle n
{\cal M} V({\cal F})\rangle_{1} = J(t), \label{20} \quad \\
V({\cal F}) = {2{\cal F}\over 1 + {\cal F}^2} , \quad
v_{M} = {\Delta l\, {\cal I}_{1}(M)\over 4\hbar\tau_{e} {\cal I}_{0}(M)}, \quad
{\cal M}\left({n\over N_{D}}\right) = { {\cal I}_{1}(\tilde{\mu}) \,
{\cal I}_{0}(M)\over {\cal I}_{1}(M)\, {\cal I}_{0}(\tilde{\mu}) },
\quad \label{21}\\
{\cal I}_{m}(s) = \int_{-\pi}^{\pi}\cos (m k)\,\ln\left( 1+e^{s-\delta+
\delta\cos k}\right)\,dk, \quad \label{22}
\end{eqnarray}
provided ${\cal F}\equiv {\cal F}_{1}$, $\delta=\Delta/(2 k_{B}T)$ and $\tilde{
\mu}\equiv\mu/(k_{B}T)$. Here $M$ (calculated graphically in Fig.~1 of 
Ref.~\cite{BEP03}) is the value of the dimensionless chemical potential $\tilde{\mu}$ at 
which (\ref{3}) holds with $n=N_{D}$. The drift velocity $v_{M} V({\cal F})$ has the 
Esaki-Tsu form with a peak velocity that becomes $v_{M}\approx \Delta l I_{1}(\delta)
/[4\hbar\tau_{e} I_{0}(\delta)]$ in the Boltzmann limit \cite{ISh87} ($I_{n}(\delta)$ 
is the modified Bessel function of the $n$th order).

To find the first-order correction in (\ref{13}), we first solve (\ref{15}) and find 
$J^{(m)}$ for $m=1$. The calculation yields the first correction to Eq.\ (\ref{20}) 
(here $'$ means differentiation with respect to $n$) \cite{BEs05}:
\begin{eqnarray}
&&\varepsilon {\partial F\over\partial t} + {e v_{M}\over l}\,
{\cal N}\left(F,{\partial F\over\partial x}\right)
=  \varepsilon\, \left\langle D\left(F,{\partial F\over\partial x},
{\partial^2 F\over \partial x^2}\right)\right\rangle_{1}
 + \langle A\rangle_{1}\, J(t) , \qquad \label{23} \\
A & = & 1 + {2 e v_{M}\over \varepsilon F_{M} l (\nu_{e}+ \nu_{i})}\,
 {1- (1+2 \tau^{2}_{e})\, {\cal F}^2\over
(1+ {\cal F}^{2})^3 }\, n {\cal M},  \label{24}\\
{\cal N} & = & \langle n V {\cal M}\rangle_{1} + \langle (A - 1)
\langle\langle n V {\cal M}\rangle_{1}\rangle_{1}\rangle_{1}
- {\Delta l \tau_{e}\over  F_{M}\hbar (\nu_{e}+
\nu_{i})}\,\left\langle {B\over 1+{\cal F}^2}\right
\rangle_{1} ,   \quad  \label{25}\\
\quad D & = & {\Delta^2 l^2 \over 8\hbar^2 (\nu_{e}+\nu_{i}) (1 + {\cal F}^{2}) }
 \left( {\partial^2\langle F\rangle_{1}\over\partial x^2} - {4\hbar v_{M}
 \tau_{e} C\over\Delta l} \right), \quad \label{26}\\
\label{27} B & = & \left\langle {4 {\cal F}_{2}n{\cal M}_{2}
\over (1+4{\cal F}^2_{2})^2} {\partial \langle F\rangle_{2}\over\partial x}
\right\rangle_{1}
+ {\cal F} \left\langle {n{\cal M}_{2} (1-4{\cal F}^2_{2})
\over (1+4{\cal F}^2_{2})^2} {\partial \langle F\rangle_{2}\over\partial x}
\right\rangle_{1} \\
&& - {4\hbar v_{M} (1+\tau_{e}^2){\cal F}(n{\cal M})'\over \Delta l
 \tau_{e}(1+{\cal F}^2)}\left\langle n{\cal M} {1-{\cal F}^{2}\over
 (1+ {\cal F}^2)^2} {\partial \langle F\rangle_{1}\over \partial x}
 \right\rangle_{1} , \nonumber\\
C & = & \left\langle { (n {\cal M}_{2})' \over 1 +  4 {\cal F}_{2}^2 }\,
{\partial^2 F\over \partial x^2}\right\rangle_{1} - 2 {\cal F}\left\langle
{ (n {\cal M}_{2})' {\cal F}_{2}\over 1 +  4 {\cal F}_{2}^2 }\,
{\partial^2 F\over \partial x^2}\right\rangle_{1} \label{28}\\
&& + {8\hbar v_{M} (1+\tau_{e}^2)
(n {\cal M})'\, {\cal F}\over \Delta l \tau_{e}\, (1 + {\cal F}^{2})}
\,\left\langle  {(n {\cal M})'{\cal F}\over 1 + {\cal F}^{2}}\,
{\partial^2 F\over \partial x^2} \right\rangle_{1}. \nonumber
\end{eqnarray}
Here ${\cal M}_{2}(n/N_{D})\equiv {\cal I}_{2}(\tilde{\mu})\, {\cal I}_{0}(M)/
[{\cal I}_{1}(M)\, {\cal I}_{0}(\tilde{\mu})]$. If the electric field and the electron 
density do not change appreciably over two SL periods, $\langle F\rangle_{j}\approx F$, 
the spatial averages can be ignored, and the {\em nonlocal}  QDDE (\ref{23}) becomes the 
{\em local} generalized DDE (GDDE) obtained from the semiclassical theory \cite{BEP03}.
The boundary conditions for the QDDE (\ref{23}) (which contains triple spatial
averages) need to be specified on the intervals $[-2l,0]$ and $[Nl,Nl+2l]$, not
just at the points $x=0$ and $x=Nl$, as in the case of the parabolic GDDE.
Similarly, the initial condition has to be defined on the extended interval
$[-2l,Nl+2l]$. For realistic values of the parameters representing a strongly coupled SL
under dc voltage bias, the numerical solution of the QDDE yields a stable self-sustained
oscillation of the current \cite{BEs05} in quantitative agreement with experiments 
\cite{sch98}. Details of the numerical procedure can be found in \cite{EB06}.

\setcounter{equation}{0}
\section{Wigner description of a two-miniband superlattice}
\label{sec:3}
We shall consider a $2\times 2$ Hamiltonian $\mathbf{H}(x,-i\partial/\partial x)$,
in which \cite{kane}
\begin{eqnarray}
&&\mathbf{H}(x,k) = [h_{0}(k)- e W(x)]\boldsymbol{\sigma}_{0}+\vec{h}(k)\cdot
\vec{\boldsymbol{\sigma}}]\nonumber\\
&& \quad\equiv\left(
\begin{array}{cc}
(\alpha+\gamma)(1-\cos kl) - e W(x) + g & - i\beta\sin kl \\
 i\beta\sin kl & (\alpha-\gamma)(1-\cos kl) - e W(x) - g 
\end{array}
\right) .\label{29}
\end{eqnarray}
Here
\begin{eqnarray}
\begin{array}{cc}
h_{0}(k)= \alpha\, (1-\cos kl), & h_{1}(k)=0, \\
h_{2}(k)=\beta\sin kl, & h_{3}(k)= \gamma\, (1-\cos kl) + g,
\end{array}  \label{30}
\end{eqnarray}
and
\begin{eqnarray}
&&\boldsymbol{\sigma}_{0}= 
\left(\begin{array}{cc}
1 & 0 \\
0 & 1
\end{array}\right),
\,\boldsymbol{\sigma}_{1}= 
\left(\begin{array}{cc}
0 & 1 \\
1 & 0
\end{array}\right),\,
\boldsymbol{\sigma}_{2}= 
\left(\begin{array}{cc}
0 & -i \\
i & 0
\end{array}\right),\,
\boldsymbol{\sigma}_{3}= 
\left(\begin{array}{cc}
1 & 0 \\
0 & -1
\end{array}\right) \label{31}
\end{eqnarray}
are the Pauli matrices.

The Hamiltonian (\ref{29}) corresponds to the simplest $2\times 2$ Kane model in which
the quadratic and linear terms $(kl)^2/2$ and $kl$ are replaced by $(1-\cos kl)$ and $\sin 
kl$, respectively. For a SL with two minibands, $2g$ is the miniband gap and $\alpha= 
(\Delta_{1}+\Delta_{2})/4$ and $\gamma = (\Delta_{1}-\Delta_{2})/4$, provided
$\Delta_{1}$ and $\Delta_{2}$ are the miniband widths. In the case of a LSL, $g=
\gamma=0$, and $h_{2}\boldsymbol{\sigma}_{2}$ corresponds to the precession term in 
the Rashba spin-orbit interaction \cite{KBB05}. The other term, the intersubband coupling, 
depends on the momentum in the $y$ direction and we have not included it here. Small 
modifications of (\ref{29}) represent a single miniband SL with dilute magnetic impurities 
in the presence of a magnetic field $B$: $g=\gamma=h_{2}=0$, and $h_{1}= \beta(B)$ 
\cite{SMP02}. As in the case of a single miniband SL, $W(x)$ is the electric potential.

The energy minibands ${\cal E}^{\pm}(k)$ are the eigenvalues of the free Hamiltonian
$\mathbf{H}_{0}(k)= h_{0}(k)\boldsymbol{\sigma}_{0}+\vec{h}(k)\cdot
\vec{\boldsymbol{\sigma}}$ and are given by
\begin{equation}
\mathcal{E}^{\pm}(k) = h_{0}(k) \pm |\vec{h}(k)|. \label{32}
\end{equation}
The corresponding spectral projections are 
\begin{equation}
\mathbf{P}^{\pm}(k) = {\boldsymbol{\sigma}_{0} \pm \vec{\nu}(k)\cdot\vec{
\boldsymbol{\sigma}}\over 2},\quad\mbox{where}\quad \vec{\nu}(k) =\vec{h}(k)/
|\vec{h}(k)|, \label{33}
\end{equation}
so that we can write
\begin{equation}
\mathbf{H}_{0}(k)= \mathcal{E}^{+}(k)\mathbf{P}^{+}(k) + \mathcal{E}^{-}(k)
\mathbf{P}^{-}(k). \label{34}
\end{equation}

We shall now write the WPBGK equations for the Wigner matrix written in terms of the 
Pauli matrices:
\begin{equation}
\mathbf{f}(x,k,t) = \sum_{i=0}^{3} f^{i}(x,k,t)\boldsymbol{\sigma}_{i} =
f^0(x,k,t)\boldsymbol{\sigma}_{0} + \vec{f}(x,k,t)\cdot\vec{
\boldsymbol{\sigma}}. \label{35}
\end{equation}
The Wigner components are real and can be related to the coefficients of the Hermitian
Wigner matrix by
\begin{eqnarray}
\begin{array}{cc}
f_{11}= f^0 + f^3, & f_{12}= f^1 - if^2, \\
f_{21}=f^1 + i f^2, & f_{22}= f^0 - f^3.
\end{array}  \label{36}
\end{eqnarray}
Hereinafter we shall use the equivalent notations
\begin{eqnarray}
f= \left(\begin{array}{c}
f^0 \\
\vec{f}
\end{array}\right) =  
\left(\begin{array}{c}
f^0 \\
f^1\\
f^2\\
f^3
\end{array}\right). \label{37}
\end{eqnarray}
The populations of the minibands with energies $\mathcal{E}^\pm$ are given by the
moments:
\begin{equation}
n^\pm(x,t) = {l\over 2\pi}\int_{-\pi/l}^{\pi/l} \left[
f^0(x,k,t)\pm \vec{\nu}(k)\cdot \vec{f}(x,k,t)\right]\, dk, \label{38}
\end{equation}
and the total electron density is $n^+ + n^-$. After some algebra, we can obtain the 
following WPBGK equations for the Wigner components
\begin{eqnarray}
&&{\partial f^0\over\partial t} + {\alpha\over\hbar}\sin kl\,\Delta^- f^0
+\vec{b}\cdot\Delta^-\vec{f} - \Theta f^0 = Q^0[f],\label{39}\\
&&{\partial\vec{f}\over\partial t} + {\alpha\over\hbar}\sin kl\,\Delta^-
\vec{f} + \vec{b}\,\Delta^- f^0 + \vec{\omega}\times\vec{f} - \Theta\vec{f} =
\vec{Q}[f], \label{40}\\
&&\varepsilon\, {\partial^2 W\over\partial x^2} = {e\over l}\, 
(n^+ + n^- -N_{D}),  \label{41}
\end{eqnarray}
whose right hand sides contain collision terms to be described later. Here
\begin{eqnarray}
&&(\Delta^\pm u)(x,k) = u(x+l/2,k)\pm u(x-l/2,k),   \label{42}\\
&&\vec{\omega} = \vec{\omega}_{0} + \vec{\omega}_{1}, \label{43}\\
&&\vec{\omega}_{0} = {2g\over\hbar}\, (0,0,1), \label{44}\\
&&\vec{\omega}_{1}= {1\over\hbar}\, (0, \beta\sin kl\,\Delta^+,2\gamma-
\gamma\cos kl\,\Delta^+), \label{45}\\
&&\vec{b} = {1\over\hbar}\, (0,\beta\cos kl,\gamma\sin kl), \label{46}\\
&&\Theta f^{i} (x,k,t)= \sum_{j=-\infty}^\infty {ejl\over i\hbar} \langle F(x,t)
\rangle_{j} e^{ijkl} f^{i}_{j}(x,t).\label{47}
\end{eqnarray}
Our collision model contains two terms: a BGK term which tries to send the miniband Wigner
function to its local equilibrium and a scattering term from the miniband with higher
energy to the lowest miniband:
\begin{eqnarray}
&& Q^0[f] = - {f^0 - \Omega^0\over\tau},  \label{48}\\
&&\vec{Q}[f] = - {\vec{f} - \vec{\Omega}\over\tau} - 
{\vec{\nu} f^0 + \vec{f}\over\tau_{\rm sc}}, \label{49}\\
&&\Omega^{0} = {\phi^+ + \phi^-\over 2}\,, \quad
\vec{\Omega}  = {\phi^+ - \phi^-\over 2}\,\vec{\nu},\label{50}\\
&&\phi^{\pm}(k;n^\pm) = {m^{*}k_{B}T\over \pi\hbar^2}\,\int_{-\infty}^\infty 
\frac{\sqrt{2}\,\Gamma^3/\pi}{\Gamma^4 +[E-{\cal E}^\pm(k)]^4}\,
\ln\left(1+ e^{{\mu^\pm - E\over k_{B}T}}\right)\, dE,\label{51}\\
&& n^\pm = {l\over 2\pi}\int_{-\pi/l}^{\pi/l} \phi^\pm(k;n^\pm)\, dk. \label{52}
\end{eqnarray}
The chemical potentials of the minibands, $\mu^+$ and $\mu^{-}$ are calculated in terms
of $n^+$ and $n^-$ respectively, by inserting (\ref{51}) in (\ref{52}) and solving the
resulting equations. Our collision model should enforce charge continuity. To check this, 
we first calculate the time derivative of $n^\pm$ using (\ref{38}) to (\ref{40}):
\begin{eqnarray}
&&{\partial n^\pm\over\partial t} + {\alpha l\Delta^-\over 2\pi\hbar}
\int_{-\pi/l}^{\pi/l} \sin kl\,(f^0\pm \vec{\nu}\cdot\vec{f})\, dk
+{l\Delta^-\over 2\pi}\int_{-\pi/l}^{\pi/l}(\vec{b}\cdot\vec{f} \pm \vec{\nu}
\cdot \vec{b} f^0)\, dk\label{53}\\
&&\quad \pm {l\Delta^-\over 2\pi}\int_{-\pi/l}^{\pi/l} \vec{\nu}\cdot
\vec{\omega}\times\vec{f}\, dk \mp {l\Delta^-\over 2\pi}\int_{-\pi/l}^{\pi/l}
\vec{\nu}\cdot\Theta\vec{f}\, dk \nonumber\\
&&\quad = {l\Delta^-\over 2\pi}\int_{-\pi/l}^{\pi/l} (Q^0[f]\pm\vec{\nu}\cdot
\vec{Q}[f])\, dk = \mp {n^+ \over\tau_{\rm sc}},  \nonumber
\end{eqnarray}
where we have employed $\int \Theta f^0 dk =0$. Then we obtain:
\begin{eqnarray}
{\partial\over\partial t}(n^+ + n^-) + \Delta^-\left[
{l\over\pi}\int_{-\pi/l}^{\pi/l} \left({\alpha\over\hbar}\sin kl\,f^0 + \vec{b}
\cdot\vec{f}\right) dk \right] = 0. \label{54}
\end{eqnarray}
Noting that $\Delta^- u(x)= l\,\partial\langle u(x)\rangle_{1}/\partial x$, we 
see that this equation corresponds to charge continuity. Differentiating in time the Poisson 
equation (\ref{41}), using (\ref{54}) in the result and integrating with respect to $x$, 
we get the following nonlocal Amp\`ere's law for the balance of current:
\begin{eqnarray}
\varepsilon {\partial F\over\partial t} + \left\langle
{el\over\pi}\int_{-\pi/l}^{\pi/l} \left({\alpha\over\hbar}\sin kl\,f^0 + \vec{b}
\cdot\vec{f}\right) dk \right\rangle_{1} = J(t). \label{55}
\end{eqnarray}
Here the space independent function $J(t)$ is the total current density. Since the Wigner
components are real, we can rewrite (\ref{55}) in the following equivalent form:
\begin{eqnarray}
\varepsilon {\partial F\over\partial t} - {2e\over\hbar}\,\left\langle
\alpha\,\mbox{Im}f^0_{1} - \beta\,\mbox{Re} f^2_{1}
+\gamma\, \mbox{Im}f^3_{1}\right\rangle_{1} = J(t). \label{56}
\end{eqnarray}

\setcounter{equation}{0}
\section{Derivation of balance equations by the Chapman-Enskog method}
\label{sec:4}
In this Section, we shall derive the reduced balance equations for our two-miniband SL
using the Chapman-Enskog method. First of all, we should decide the order of magnitude
of the terms in the WPBGK equations (\ref{39}) and (\ref{40}) in the hyperbolic limit.
Recall that in this limit, the collision frequency $1/\tau$ and the Bloch frequency 
$eF_{M}l/\hbar$ are of the same order, about 10 THz for the SL of Section \ref{sec:2}. 
Typically, $2g/\hbar$ is of the same order, so that the term containing $\vec{\omega}_{0}$ 
should also balance the BGK collision term. What about the other terms?

The scattering time $\tau_{\rm sc}$ is  much longer than the collision time $\tau$, 
and we shall consider $\tau/\tau_{\rm sc}= O(\lambda)\ll 1$. Moreover, the gap 
energy is typically much larger than the miniband widths or the spin-orbit coefficient and
a rich dominant balance is obtained by assuming that $\beta/g$ and $\gamma/g$ are of order 
$\lambda$. Then we can expand the unit vector $\vec{\nu}$ as follows:
\begin{eqnarray}
\vec{\nu} = (0,0,1) + {\lambda\beta\over g}\,\sin kl\, (0,1,0)
- \lambda^2\left[{\beta\gamma\over g^2}\,\sin kl (1-\cos kl)\, (0,1,0)\right.
\label{57}\\
\left. + {\beta^2\sin^2kl\over 2 g^2}\, (0,0,1)\right] + O(\lambda^3). \nonumber
\end{eqnarray}
In this expansion, we have inserted the book-keeping parameter $\lambda$ which is set 
equal to 1 at the end of our calculations (cf.\ Section \ref{sec:2}). From (\ref{39}) and 
(\ref{40}), we can write the scaled WPBGK equations as follows:
\begin{eqnarray}
\mathbb{L} f -\Omega = - \lambda\,\left(\tau\,{\partial f\over\partial t}+
\Lambda f\right). \label{58}
\end{eqnarray}
Here the operators $\mathbb{L}$ and $\Lambda$ are defined by
\begin{eqnarray}
&&\mathbb{L} f= f - \tau\, \Theta f + \delta_{1} \left( \begin{array}{c}
0\\
- f^2\\
f^{1}\\
0
\end{array}\right),  \label{59}\\
&& \Lambda f = \delta_{2}\, \left( \begin{array}{c}
0\\
\vec{f} + \vec{\nu} f^0
\end{array}\right) +{\alpha\tau\over\hbar}\sin kl\,\Delta^- f
+\Delta^- \left( \begin{array}{c}
\tau \vec{b}\cdot\vec{f}\\
\tau\, \vec{b}\, f^0 \end{array}\right)
+\left( \begin{array}{c}
0\\
\tau\, \vec{\omega}_{1}\times\vec{f} \end{array}\right),
  \label{60}
\end{eqnarray}
where
\begin{equation}
\delta_{1}= {2g\tau\over\hbar},\quad \delta_{2}= {\tau\over\tau_{\rm sc}}.
\label{61}
\end{equation}
The expansion of $\vec{\nu}$ in powers of $\lambda$ gives rise to a similar expansion of 
$\Omega$ and $\Lambda$. 

To derive the reduced balance equations, we use the following Chapman-Enskog ansatz:
\begin{eqnarray}
&& f(x,k,t;\lambda) = f^{(0)}(k;n^+,n^-,F) + \sum_{m=1}^{\infty} 
f^{(m)}(k;n^+,n^-,F)\, \lambda^{m} ,    \label{62}\\
&& \varepsilon {\partial F\over\partial t} + \sum_{m=0}^{\infty}  
J_{m}(n^+,n^-,F)\, \lambda^{m} = J(t),  \label{63}\\
&&{\partial n^\pm\over\partial t} = \sum_{m=0}^{\infty}  
A^{\pm}_{m}(n^+,n^-,F)\, \lambda^{m}.  \label{64}
\end{eqnarray}
The functions $A_{m}^\pm$ and $J_{m}$ are related through the Poisson equation 
(\ref{41}), so that
\begin{eqnarray}
A^{+}_{m}+ A^-_{m} = - {l\over e} \, {\partial J_{m}\over\partial x}.  
\label{65}
\end{eqnarray}
Inserting (\ref{62}) to (\ref{64}) into (\ref{58}), we get
\begin{eqnarray}
&& \mathbb{L} f^{(0)} = \Omega_{0}, \label{66}\\
&& \mathbb{L} f^{(1)} = \Omega_{1} - \left.\tau\,
{\partial f^{(0)}\over\partial t}\right|_{0} - \Lambda_{0} f^{(0)},\label{67}\\
&&\mathbb{L} f^{(2)} = \Omega_{2} - \left.\tau\, {\partial f^{(1)}\over\partial 
t}\right|_{0} - \Lambda_{0} f^{(1)} - \left.\tau\,{\partial f^{(0)}\over\partial 
t}\right|_{1} - \Lambda_{1} f^{(0)}, \label{68}
\end{eqnarray}
and so on. The subscripts 0 and 1 in the right hand side of these equations mean that we
replace $\varepsilon\,\partial F/\partial t|_{m}= J \delta_{0m}-J_{m}$, 
$\partial n^\pm/\partial t|_{m}=A^\pm_{m}$. Moreover, inserting (\ref{57}) and 
(\ref{62}) into (\ref{38}) yields the following compatibility conditions:
\begin{eqnarray}
&& f^{(1)\,0}_{0}=0, \quad f^{(1)\, 3}_{0} = {\beta\over g}\,\mbox{Im}
f^{(0)\, 2}_{1},\label{69}\\
&& f^{(2)\, 0}_{0} = 0, \label{70}\\
&& f^{(2)\, 3}_{0} = {\beta\over g}\,\mbox{Im} f^{(1)\, 2}_{1} + {\beta^2\over 
4g^2}\, (f^{(0)\, 3}_{0}- \mbox{Re} f^{(0)\, 3}_{2}) -  {\beta\gamma\over g^2}
\, \mbox{Im}\left(f^{(0)\, 2}_{1}- {f^{(0)\, 2}_{2}\over 2}\right) ,\nonumber
\end{eqnarray}
etc. 

To solve (\ref{66}) for $f^{(0)}\equiv\varphi$, we first note that 
\begin{eqnarray}
&&-\tau\, \Theta\varphi =  \sum_{j=-\infty}^\infty i \vartheta_{j} 
\varphi_{j} e^{ijkl},  \label{71}\\
&& \vartheta_{j} \equiv {\tau ejl\over\hbar}\,\langle F\rangle_{j}. \label{72}
\end{eqnarray}
Then (\ref{66}), (\ref{50}) and (\ref{57}) yield
\begin{eqnarray}
\varphi_{j}^0 =  {\phi^+_{j} + \phi^-_{j}\over 2}\, {1- i \vartheta_{j}\over 1
+ \vartheta_{j}^2},\quad \varphi_{j}^1 = \varphi_{j}^2 = 0,  \quad 
\varphi_{j}^3 =  {\phi^+_{j} - \phi^-_{j}\over 2}\, {1- i \vartheta_{j}\over 1
+ \vartheta_{j}^2},   \label{73}
\end{eqnarray}
where we have used that the Fourier coefficients
\begin{eqnarray}
\phi_{j}^\pm =  {l\over \pi}\, \int_{0}^{\pi/l}\cos(jkl)\,\phi^\pm\, dk, \label{74}
\end{eqnarray}
are real because $\phi^\pm$ are even functions of $k$. Similarly, the solution of (\ref{67})
is $f^{(1)}\equiv \psi$ with
\begin{eqnarray}
&& \psi_{j}^m =  r^m_{j}\, {1- i \vartheta_{j}\over 1 + \vartheta_{j}^2}
\quad (m=0,3),\nonumber\\ 
&& \psi_{j}^1 = {(1+ i \vartheta_{j})\, r_{j}^1+ \delta_{1}\,
r_{j}^2\over (1+ i \vartheta_{j})^2 +\delta_{1}^2},\label{75}\\
&&\psi_{j}^2 = {(1+ i \vartheta_{j})\, r_{j}^2 -\delta_{1}\, 
r^1_{j}\over (1+ i \vartheta_{j})^2 +\delta_{1}^2}.\nonumber
\end{eqnarray}
Here $r$ is the right hand side of (\ref{67}). The balance equations can be found in two 
ways. We can calculate $A_{m}^\pm$ for $m=0,1$ by using the compatibility conditions
(\ref{69}) and (\ref{70}) in Equations (\ref{67}) and (\ref{68}), respectively. More
simply, we can insert the solutions (\ref{73}) and (\ref{75}) in the balance equations
(\ref{53}) and in the Amp\`ere's law (\ref{55}). The result is: 
\begin{eqnarray}
&& {\partial n^\pm\over\partial t} + \Delta^- D_{\pm}(n^+,n^-,F)=
\pm R(n^+,n^-,F),\label{76}\\ 
&& \varepsilon\,{\partial F\over\partial t}+{e\over\hbar}\,\left\langle
[\alpha\, (\phi^+_{1}+\phi^-_{1}) + \gamma\, (\phi^+_{1}-\phi^-_{1})]\, {
 \vartheta_{1}\over 1+ \vartheta_{1}^2}\right\rangle_{1} \label{77}\\
 &&\quad +  {2e\over\hbar}\, [\beta\mbox{Re}\langle\psi^2_{1}\rangle_{1}
 - \alpha\,\mbox{Im}\langle\psi^0_{1}\rangle_{1} - \gamma\,\mbox{Im}
 \langle\psi^3_{1}\rangle_{1}] = J ,\nonumber\\
&&D_{\pm} = {\alpha\pm\gamma\over\hbar}\, \left[{\phi_{1}^\pm 
\vartheta_{1}\over 1+ \vartheta_{1}^2} - \mbox{Im}(\psi^0_{1}\pm\psi^3_{1})
\right]+ {\beta\over\hbar}\, \mbox{Re}\psi^2_{1}\pm {\beta^2\vartheta_{2}\over
4 g\hbar}\, {\phi_{2}^+ + \phi_{2}^-\over 1 + \vartheta_{2}^2},\label{78}\\
&& R = -{\delta_{2}n^+\over\tau} - {\beta^2\vartheta_{2}^2 (\phi^+_{2}
-\phi^-_{2})
\over 8 g^2\tau (1+\vartheta_{2}^2)} +{\beta\over g\tau}\,\vartheta_{1}
\mbox{Re}\psi^2_{1} + {\beta\over\hbar}\, (2-\Delta^+)\mbox{Im}
\psi^1_{1}.\label{79}
\end{eqnarray}
Appendix \ref{appA} justifies this second and more direct method by showing that 
equivalent expressions are obtained from the compatibility conditions. Note that Eq.\ 
(\ref{77}) can be obtained from (\ref{76}) and the Poisson equation.

\setcounter{equation}{0}
\section{Spintronics: Quantum drift-diffusion equations for a lateral superlattice with 
Rashba spin-orbit interaction}
\label{sec:5}
In the simpler case of a LSL with the precession term of Rashba spin-orbit interaction
(but no intersubband coupling), we can obtain explicit rate equations for $n^\pm$ by
means of the Chapman-Enskog method. In the Hamiltonian (\ref{29}), we have $\gamma=
g=0$, so that $h_{3}=0$ and $\vec{\nu}=(0,1,0)$. However, the Fermi-Dirac distribution
is different from (\ref{4.1}) for a LSL. We have to replace $E_{n}$ instead of 
$\hbar^2k_{z}^2/(2m^*)$, sum over $n$ for all populated QW energy levels and integrate 
over $k_{y}$ only. Provided only $E_{1}$ is populated, we obtain the following expression
instead of (\ref{51}):
\begin{eqnarray}
\phi^\pm(k;n^\pm) = \int_{-\infty}^\infty {D_{\Gamma}\left(E-{\cal E}^\pm(k)
-E_{1}\right)\over 1+\exp\left({E-\mu^\pm\over k_{B}T}\right)}\, dE, 
\label{80}
\end{eqnarray} 
where the broadened density of states is
\begin{eqnarray}
\quad D_{\Gamma}(E) = {1\over 2\pi L_{z}}\int_{-\infty}^\infty dk_{y}
\delta_{\Gamma}\left( {\hbar^2k_{y}^2\over 2 m^*} - E\right) = 
{\sqrt{2m^*}\over 2\pi\hbar L_{z}}\int_{0}^\infty dE_{y}\,
{\delta_{\Gamma}( E_{y} - E)\over\sqrt{E_{y}}}. \label{81}
\end{eqnarray} 
Note that (\ref{81}) becomes the 1D density of states $D(E)= \sqrt{2m^*}\theta(E)/(2
\pi\hbar L_{z}\sqrt{E})$ as $\Gamma\to 0+$. We have not included a factor 2 in 
(\ref{81}) because all the electrons in each of the minibands (with energies ${\cal E}^\pm
(k)$) have the same spin. Inserting (\ref{4.3}) in (\ref{81}) and using the residue theorem
to evaluate the integral, we obtain
\begin{eqnarray}
\quad\quad D_{\Gamma}(E) &=& {\sqrt{m^*} \over 4\pi\hbar L_{z}}\label{82}\\
&\times&\left[{\sqrt{\sqrt{E^2+\sqrt{2}\Gamma E+\Gamma^2}+E+
{\Gamma\over\sqrt{2}}}- \sqrt{\sqrt{E^2+\sqrt{2}\Gamma E+\Gamma^2} -E- 
{\Gamma\over\sqrt{2}}}\over \sqrt{E^2+\sqrt{2}\Gamma E+\Gamma^2}}\right.
\nonumber \\
&+&\left. {\sqrt{
\sqrt{E^2-\sqrt{2}\Gamma E+\Gamma^2} +E- {\Gamma\over\sqrt{2}}}+
\sqrt{\sqrt{E^2-\sqrt{2}\Gamma E+\Gamma^2}-E +{\Gamma\over\sqrt{2}}}
\over \sqrt{E^2-\sqrt{2}\Gamma E+\Gamma^2}}\right].\nonumber
\end{eqnarray} 
As $E\to +\infty$, $D_{\Gamma}(E)\sim \sqrt{2m^*}/(2\pi\hbar L_{z}\sqrt{E})$,
whereas $D_{\Gamma}(E) = O(|E|^{-5/2})$ as $E\to -\infty$. Then the convolution 
integral (\ref{80}) is convergent. 

In the present case, minibands correspond to electrons with spin up or down which have
different energy. Scattering between minibands is the same as in (\ref{49}), $-(\vec{\nu}
f^0+\vec{f})/\tau_{\rm sc}$ which yields $\partial n^\pm/\partial t +\ldots = \mp 
n^\pm/\tau_{\rm sc}$ in (\ref{53}), only if the chemical potential of the miniband with
lowest energy, $\mu^-$, is less than the minimum energy of the other miniband, ${\cal 
E}^+_{\rm min}= $min$_{k}{\cal E}^+(k)$. Otherwise ($\mu^->{\cal E}^+_{\rm 
min}$), the scattering term should be $-2\vec{f}/\tau_{\rm sc}$, which yields $\partial 
n^\pm/\partial t +\ldots = \mp (n^+ - n^-)/\tau_{\rm sc}$ in (\ref{53}), thereby trying 
to equalize $n^+$ and $n^-$; cf.\ Ref.~\cite{SMP02}.

Now we shall derive the balance equations in the hyperbolic limit using the Chapman-Enskog 
method as in Section \ref{sec:4}. In the scaled WPBGK equations (\ref{58}), the operators 
$\mathbb{L}$ and $\Lambda$ are
\begin{eqnarray}
\mathbb{L} f &=& f - \tau\, \Theta f,  \label{5.2}\\
\Lambda f &=& \delta_{2} \left( \begin{array}{c}
0\\
2\vec{f} + (\vec{\nu} f^0-\vec{f})\, \theta({\cal E}^+_{\rm min}-\mu^-)
\end{array}\right) +{\alpha\tau\over\hbar}\sin kl\,\Delta^- f\label{5.3}\\
&+& {\beta\tau\over\hbar}\,\cos kl\, \Delta^- \left( \begin{array}{c}
f^2\\
0\\
f^0\\
0 \end{array}\right)
+ {\beta\tau\over\hbar}\,\sin kl\, \Delta^+ \left( \begin{array}{c}
0\\
f^3\\
0\\
- f^1 \end{array}\right),
 \nonumber
\end{eqnarray}
where $\delta_{2}$ is given by (\ref{61}), $\theta(x)$ is the Heaviside unit step function 
and $\Omega^0=(\phi^+ + \phi^-)/2$, $\vec{\Omega}=(0,1,0)\, (\phi^+ - \phi^-)/2$. 
The hierarchy of equations (\ref{66}) - (\ref{68}) is simply
\begin{eqnarray}
&& \mathbb{L} f^{(0)} = \Omega, \label{5.9}\\
&& \mathbb{L} f^{(1)} = - \left.\tau\,
{\partial f^{(0)}\over\partial t}\right|_{0} - \Lambda f^{(0)},\label{5.10}\\
&&\mathbb{L} f^{(2)} = - \left.\tau\, {\partial f^{(1)}\over\partial 
t}\right|_{0} - \Lambda f^{(1)} - \left.\tau\,{\partial f^{(0)}\over\partial 
t}\right|_{1}, \label{5.11}
\end{eqnarray}
and so on. The compatibility and solvability conditions are:
\begin{eqnarray}
f^{(m)\, 0}_{0} = f^{(m)\, 2}_{0} = 0\quad \Longrightarrow (\mathbb{L} 
f^{(m)\, 0})_{0} = (\mathbb{L} f^{(m)\, 2})_{0} = 0,\quad m\geq 1.  
\label{5.13}
\end{eqnarray}

The solution $f^{(0)}\equiv\varphi$ of (\ref{5.9}) is
\begin{eqnarray}
\varphi_{j}^0 =  {\phi^+_{j} + \phi^-_{j}\over 2}\, {1- i \vartheta_{j}\over 1
+ \vartheta_{j}^2},\quad \varphi_{j}^1 = \varphi_{j}^3 = 0,  \quad 
\varphi_{j}^2 =  {\phi^+_{j} - \phi^-_{j}\over 2}\, {1- i \vartheta_{j}\over 1
+ \vartheta_{j}^2},   \label{5.16}
\end{eqnarray}
where we have used that the Fourier coefficients $\phi^\pm_{j}$ are real because $\phi^\pm$
are even functions of $k$. Similarly, the solution of (\ref{5.10})
is $f^{(1)}\equiv \psi$ with
\begin{eqnarray}
\psi_{j}^m =  r^m_{j}\, {1- i \vartheta_{j}\over 1 + \vartheta_{j}^2}
\quad (m=0,2),\quad \psi_{j}^1 = \psi_{j}^3 = 0.\label{5.18}
\end{eqnarray}
Here $r$ is the right hand side of (\ref{5.10}). The balance equations can be found in two 
ways. We can calculate $A_{m}^\pm$ for $m=0,1$ by using the solvability conditions
(\ref{5.13}) in Equations (\ref{5.10}) and (\ref{5.11}), respectively. More
simply, we can insert the solutions (\ref{5.16}) and (\ref{5.18}) in the balance equations
(\ref{53}) and in the Amp\`ere's law (\ref{55}). In both cases, the result is: 
\begin{eqnarray}
&& {\partial n^\pm\over\partial t} + \Delta^- D_{\pm}(n^+,n^-,F)=
\mp R(n^+,n^-,F),\label{5.19}\\
&& \varepsilon\,{\partial F\over\partial t}+ e\,\langle D_{+}+ D_{-}
\rangle_{1} = J, \label{5.20}\\
&& D_{\pm} = - {\alpha\over\hbar}\,\Delta^-\mbox{Im}(\varphi^{0}_{1}\pm 
\varphi^2_{1} +\psi^0_{1}\pm \psi^{2}_{1})Ê\pm {\beta\over\hbar}\,\Delta^-
\mbox{Re} (\varphi^{0}_{1}\pm\varphi^2_{1} +\psi^0_{1}\pm \psi^{2}_{1}),
\label{5.21}\\
&& R = {n^+ - n^-\,\theta(\mu^- -{\cal E}^+_{\rm min})\over\tau_{sc}}. 
\label{5.22}
\end{eqnarray}
A straightforward calculation of (\ref{5.21}) yields
\begin{eqnarray}
&& D_{\pm} =  {(\alpha\vartheta_{1}\pm\beta)\phi_{1}^\pm\over\hbar\,
(1+\vartheta_{1}^2)} \mp {\tau\, (\phi_{1}^+ - \phi_{1}^-)\, [2\alpha
\vartheta_{1}\pm\beta (1-\vartheta_{1}^2)]\over 2\hbar\tau_{\rm sc}
(1+\vartheta_{1}^2)^2} \label{5.23}\\
&&\quad + {[2\alpha\vartheta_{1}\pm\beta (1-\vartheta_{1}^2)]\alpha\tau\over
\hbar^2 (1+\vartheta_{1}^2)^2} {\partial\phi_{1}^\pm\over\partial n^\pm}
\left[\Delta^- \left({\alpha\vartheta_{1}\pm\beta\over\hbar\, (1+
\vartheta_{1}^2)}\phi_{1}^\pm \right) \pm {\hbar\over\alpha\tau_{sc}} 
(n^+ - n^-)\right]\nonumber\\
&& \quad + {\alpha\, (3\vartheta_{1}^2-1)\pm\beta\vartheta_{1}(3-\vartheta_{1}^2)
\over\hbar (1+\vartheta_{1}^2)^3}\, {l\tau^2\phi_{1}^\pm\over\hbar
\varepsilon}\left({J\over e}-\left\langle\left\langle{\alpha\, (\phi_{1}^+ + 
\phi_{1}^-)\vartheta_{1}\over\hbar (1+\vartheta_{1}^2)}\right\rangle_{1}
\right\rangle_{1} \right.\nonumber\\
&& \quad\left. - \left\langle\left\langle{\beta\, (\phi_{1}^+ - \phi_{1}^-)
\over\hbar (1+\vartheta_{1}^2)}\right\rangle_{1}\right\rangle_{1}\right)
- {(\alpha^2+\beta^2)\tau\over 2\hbar^2 (1+\vartheta_{1}^2)}\,\Delta^- n^\pm
\nonumber\\
&& \quad + {\tau\over 2\hbar^2 (1+\vartheta_{1}^2)}\,\left[(\alpha^2-\beta^2\mp 2\alpha
\beta\vartheta_{1})\,\Delta^-\left({\phi_{2}^\pm\over 1+\vartheta_{2}^2}
\right)\right.
\nonumber\\
&& \quad\left. + [(\beta^2-\alpha^2)\vartheta_{1}\mp 2\alpha\beta]\,\Delta^-
\left({\vartheta_{2}\phi_{2}^\pm\over 1+\vartheta_{2}^2}\right)\right].
\nonumber
\end{eqnarray}

We have numerically solved the system of equations (\ref{5.19}) - (\ref{5.23}), with the 
following boundary conditions in the interval $-2l\leq x\leq 0$:
\begin{eqnarray}
&&\varepsilon\,\frac{\partial F}{\partial t} + \sigma\, F = J, \label{bcF}\\ 
&& n^+ = n^- =\frac{N_{D}}{2}, \label{bcn}
\end{eqnarray}
whereas in the collector $Nl\leq x\leq N\, (l+2)$, (\ref{bcF}) and 
\begin{eqnarray}
\frac{\partial n^\pm}{\partial x} = 0 \label{bccoll}
\end{eqnarray}
hold. We have used the following values of the parameters: $\alpha= \Delta_{1}/2=8$ meV, 
$\beta= 2.63$ meV, $d_{W}=3.1$ nm, $d_{B}=1.96$ nm, $l=d_{W}+d_{B}= 5.06$ nm, 
$L_{z}=3.1$ nm, $T=5$ K, $\tau= 5.56\times 10^{-14}$ s, $\tau_{\rm sc}=5.56\times 
10^{-13}$ s, $N_{D}=4.048\times 10^{10}$ cm$^{-2}$, $m^* = (0.067 d_{W}+ 0.15 
d_{B}) m_{0}/l$, $V=3$ V, $N=110$. We have used a large conductivity of the injecting 
contact $\sigma= 11.78\,\Omega^{-1}$m$^{-1}$. With these values, we select the following 
units to present graphically our results: $F_{M}=\hbar/(el\tau)=23.417$ kV/cm, $x_{0}=
\varepsilon F_{M}l/(eN_{D}) =19.4$ nm, $t_{0}=\hbar/\alpha = 0.082$ ps, $J_{0} = 
\alpha eN_{D}/(2\hbar)=3.94\times 10^4$ A/cm$^2$. 

\begin{figure}
\begin{center}
\includegraphics[width=6.cm,angle=0]{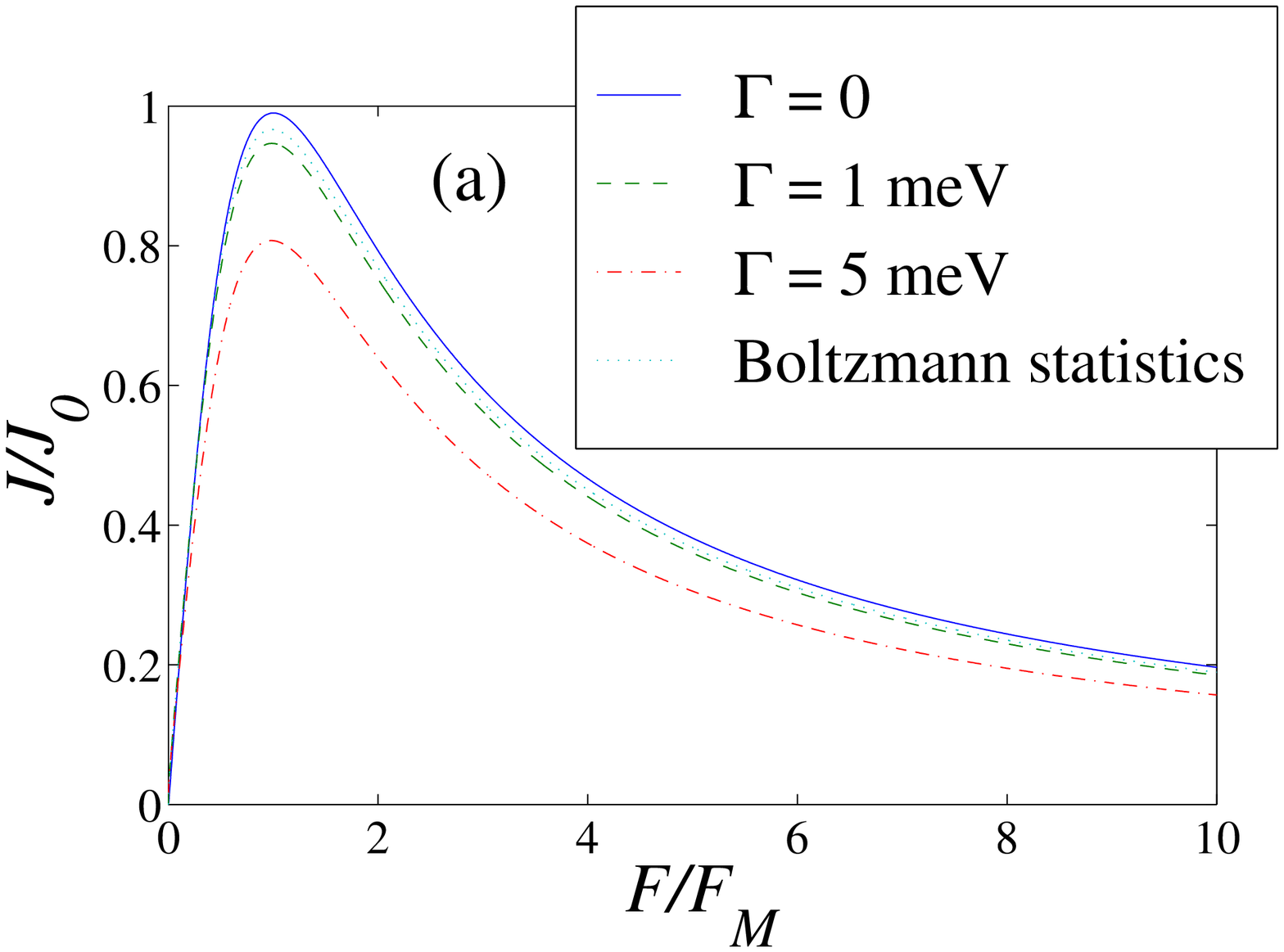}
\includegraphics[width=6.cm,angle=0]{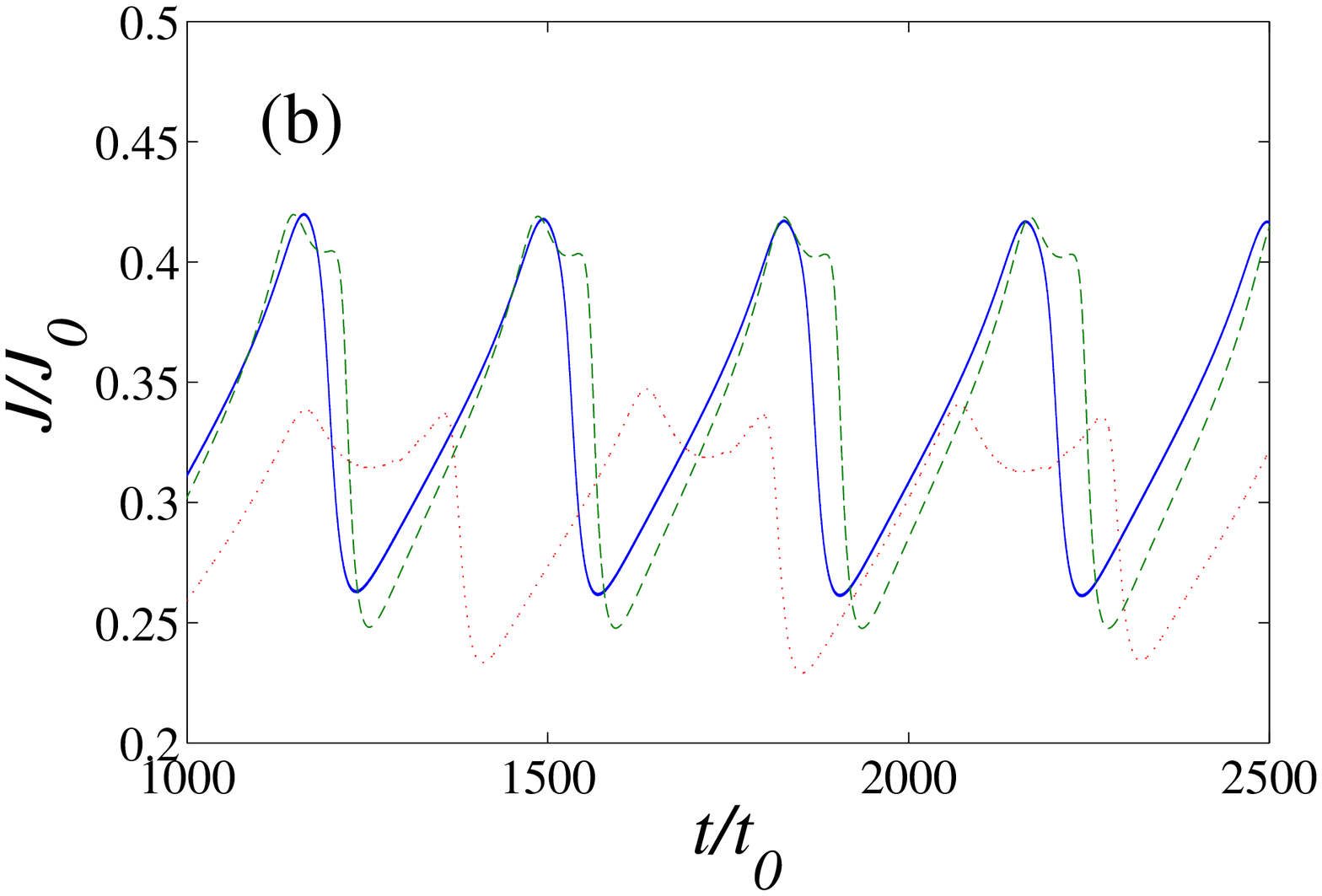}
\includegraphics[width=6.cm,angle=0]{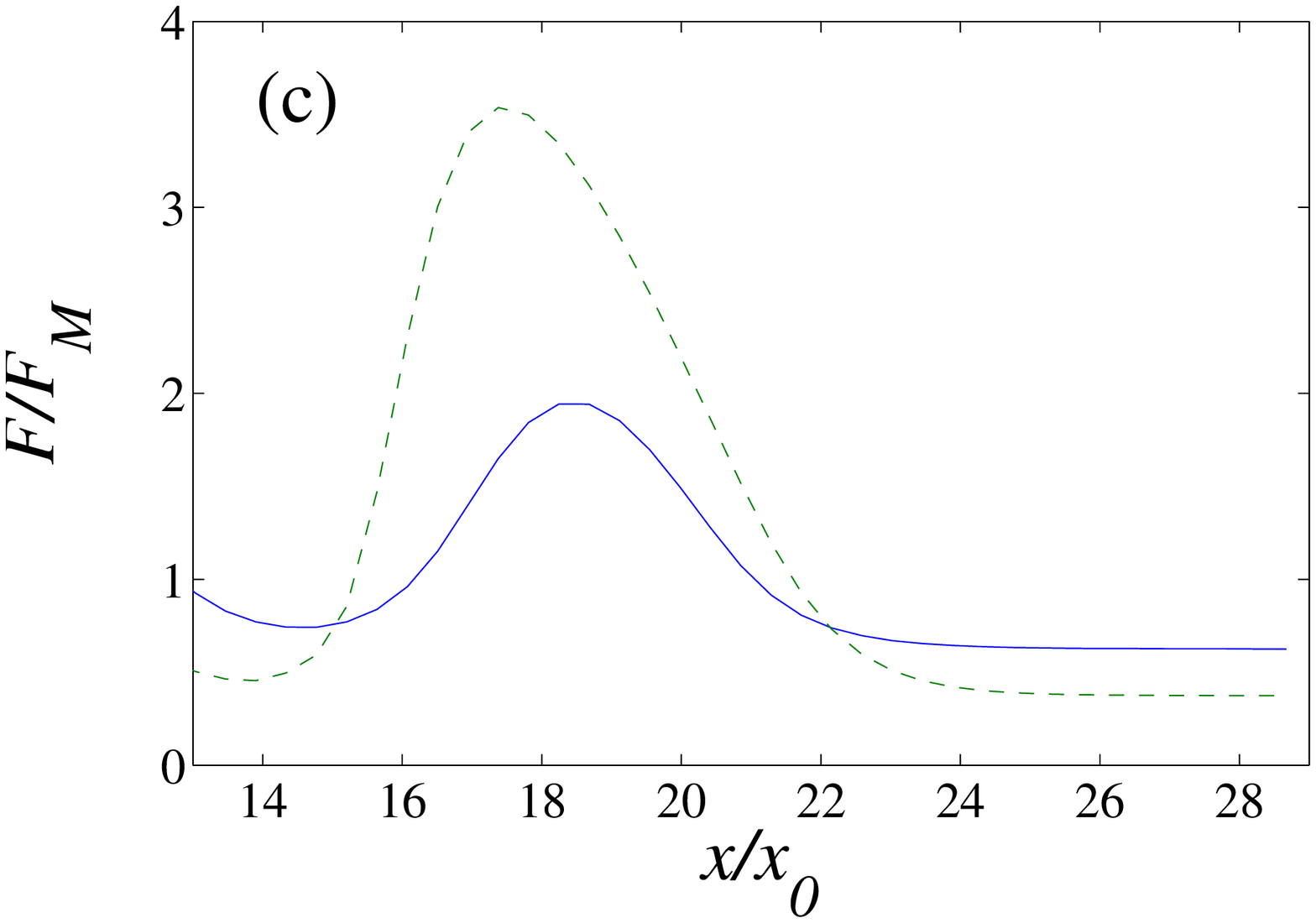}
\includegraphics[width=6.cm,angle=0]{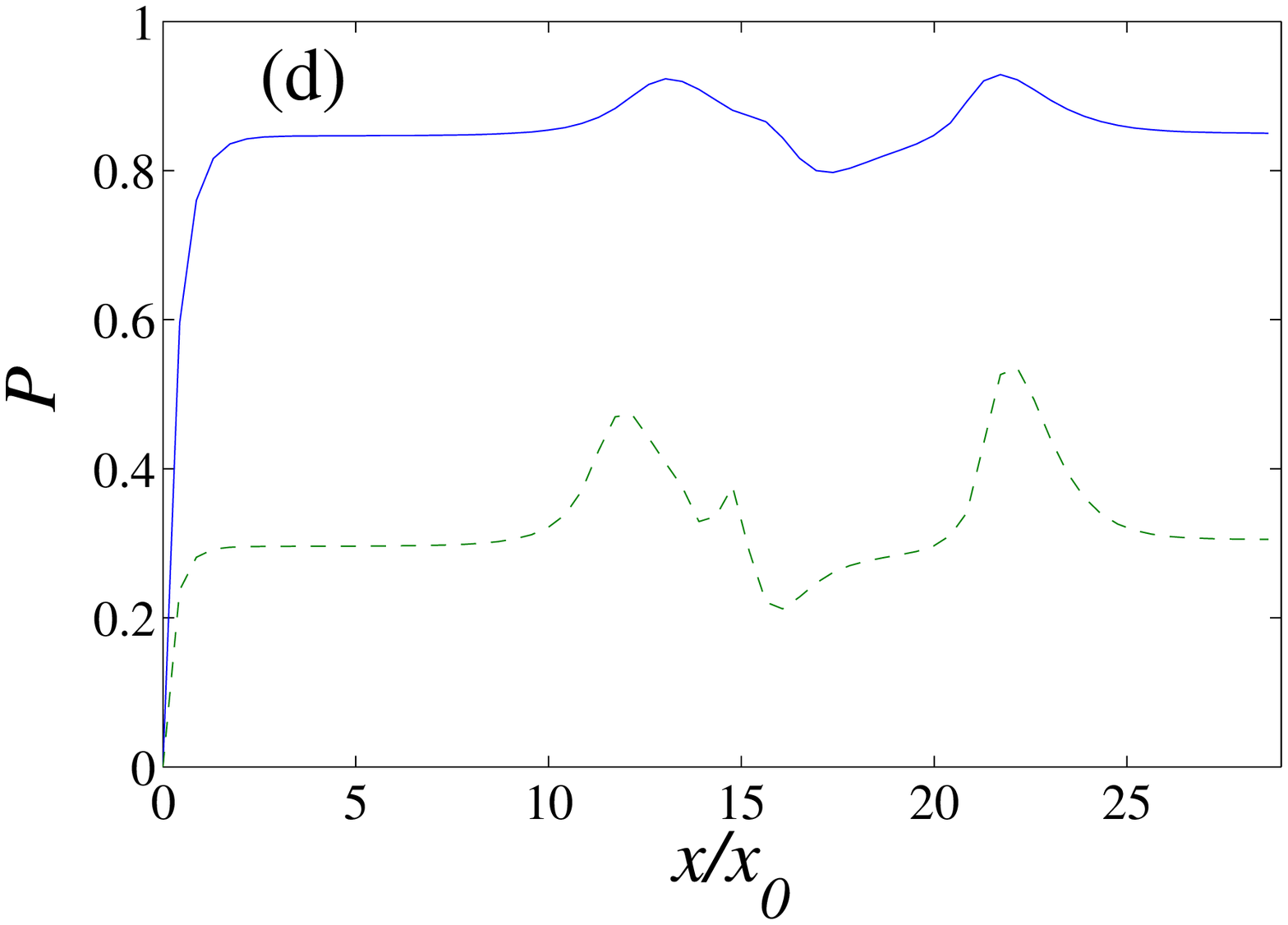}
\vspace{0.2cm}
\caption{(a) Electron current vs field in a spatially uniform stationary state for different
values of the broadening $\Gamma$ using the Fermi-Dirac distribution and for the Boltzmann 
distribution without broadening. (b) Total current density vs time, and the (c) electric field 
and (d) spin polarization profiles during current self-oscillations for $\Gamma=0$ (solid line) 
and 1 meV (dashed line). Parameter values are $N=110$, $N_{D}= 4.048\times 10^{10}$ 
cm$^{-2}$, $d_{B}= 1.96$ nm, $L_{z}=d_{W}=3.1$ nm, $l= 5.06$ nm, $\tau= 0.0556$ 
ps, $\tau_{\rm sc}= 0.556$ ps, $V=3$ V, $\sigma= 11.78\, \Omega^{-1}$m$^{-1}$ $T=5$ 
K,m$\alpha= 8$ meV, $\beta= 2.63$ meV. With these values, $\Delta_{1} = 16$ meV, 
$x_{0} = 19.4$ nm, $t_{0} = 0.082$ ps, $J_{0} = 3.94\times 10^4$ A/cm$^2$. }
\label{fig3}
\end{center}
\end{figure}

Fig.~\ref{fig3}(b) - (d) illustrates the resulting stable self-sustained current oscillations. 
They are due to the periodic formation of a pulse of the electric field at the cathode $x=0$ 
and its motion through the LSL. Fig.~\ref{fig3}(b) depicts the pulse when it is far from 
the contacts and the corresponding spin polarization is shown in Fig.~\ref{fig3}(d). 
It is interesting to consider the influence of the broadening $\Gamma$ and the Fermi-Dirac
statistics on the oscillations. At high temperatures, Boltzmann statistics and a semiclassical 
approximation should provide a good description. The semiclassical approximation is 
equivalent to dropping all spatial averages in our previous formulas. Since $x_{0}\gg l$,
the effect of dropping spatial averages should be rather small. Using Boltzmann statistics 
yields explicit formulas for $\mu^\pm$ in terms of
$n^\pm$. In fact, we only have to replace $e^{(\mu^\pm-E)/(k_{B}T)}$ instead of the 3D 
Fermi distribution $[1+e^{(E-\mu^\pm)/(k_{B}T)}]^{-1}$ in Eq.\ (\ref{80}). Using
the relation (\ref{52}) between $n^\pm$ and $\phi^\pm$, we obtain
\begin{eqnarray}
\quad\phi^\pm = n^\pm\, {\pi\,\exp\left({\alpha\,\cos kl\mp \beta\, |\sin kl|
\over k_{B}T}\right)\over\int_{0}^\pi dK\,\exp\left({\alpha\,\cos K\mp 
\beta\,\sin K\over k_{B}T}\right)}, \label{5.24}
\end{eqnarray}
and therefore, 
\begin{eqnarray}
\quad\phi^\pm_{j} = n^\pm\,{\int_{0}^\pi dK\, \cos(jK)\, \exp\left({\alpha\cos 
K\mp\beta\,\sin K\over k_{B}T}\right)\over \int_{0}^\pi dK\,\exp\left(
{\alpha\cos K\mp\beta\,\sin K\over k_{B}T}\right)},  \label{5.25}
\end{eqnarray}
for $j=0,1,\ldots$ Similar relations hold for the case of a SL with Boltzmann statistics in 
the tight-binding approximation. 

The results are shown in Fig.~\ref{fig3}. Fig.~\ref{fig3}(a) depicts the relation between
electron current and field for a spatially uniform stationary solution with $n^\pm=N_{D}/2$.
We observe that all curves are similar. However the curves for $\Gamma=0$ and $\Gamma=1$ 
meV are close while the curve for $\Gamma=5$ meV has dropped noticeably. The shapes of
$J(t)$ for $\Gamma=0$ and $\Gamma=1$ meV in Fig.~\ref{fig3}(b) are close and quite 
different from that for $\Gamma=5$ meV. If we look at the corresponding field profiles in 
Fig.~\ref{fig3}(c) and (d), for $\Gamma=0$ and $\Gamma=1$ meV the oscillations of the
current are caused by the periodic nucleation of a pulse of the electric field at $x=0$ and its
motion towards the end of the LSL. The pulse far from the contacts shown in Fig.~\ref{fig3}(c)
is larger in the case of $\Gamma=0$ than for $\Gamma=1$ meV.  In the case of $\Gamma=5$
meV (not shown), the pulse created at $x=0$ becomes attenuated and it disappears before 
arriving at $x=Nl$. This seems to indicate that the lowest voltage at which there exist stable 
self-sustained current oscillations is an increasing function of $\Gamma$: If we fix the voltage
at 3 V and increase $\Gamma$, the critical voltage threshold to have stable oscillations
approaches our fixed voltage of 3 V. Then the observed oscillations are smaller and the
field profiles correspond to waves that vanish before reaching the end of the device, as it also
occurs in models of the Gunn effect in bulk semiconductors \cite{onset}.

\setcounter{equation}{0}
\section{Conclusions}
\label{sec:6}
We have presented a Wigner-Poisson-BGK system of equations 
with a collision broadened local Fermi-Dirac distribution for strongly coupled SLs having only one populated miniband. 
In the hyperbolic limit in which the collision
and Bloch frequencies are of the same order and dominate all other frequencies, the Chapman-Enskog
perturbation method yields a quantum drift-diffusion equation for the field. Numerical solutions of this equation
exhibit self-sustained oscillations of the current due to recycling and motion of charge dipole domains \cite{BEs05}.

For strongly coupled SLs having two populated minibands, we have introduced a periodic 
version of the Kane Hamiltonian and derived the corresponding Wigner-Poisson-BGK
system of equations. The collision model comprises two terms, a BGK term trying to bring
the Wigner matrix closer to a broadened Fermi-Dirac local equilibrium at each miniband,
and a scattering term that brings down electrons from the upper to the lower miniband. 
By using the Chapman-Enskog method, we have derived quantum drift-diffusion equations
for the miniband populations which contain generation-recombination terms. As it should be,
the recombination terms vanish if there is no inter-miniband scattering and the off-diagonal 
terms in the Hamiltonian are zero. These terms may represent a Rashba spin-orbit 
interaction for a lateral superlattice. For a lateral superlattice under dc voltage bias in the growth direction, numerical solutions of the
corresponding quantum drift-diffusion equations show self-sustained
current oscillations due to periodic recycling and motion of electric field pulses. The periodic changes of the spin polarization and
spin polarized current indicate that this system acts as a spin oscillator.

\acknowledgments 
This 
research was supported by the Spanish MECD grant MAT2005-05730-C02-01.  

\appendix
\renewcommand{\theequation}{A.\arabic{equation}}
\setcounter{equation}{0}
\section{Balance equations from compatibility conditions}
\label{appA}
We know that $\varphi^1=\varphi^2=0$ from (\ref{67}). Then the compatibility 
conditions (\ref{69}) and (\ref{70}) become
\begin{eqnarray}
&& \psi^{0}_{0}=0, \quad \psi^{3}_{0} = 0,\label{a1}\\
&& f^{(2)\, 0}_{0} = 0, \quad f^{(2)\, 3}_{0} = {\beta\over g}\,\mbox{Im} 
\psi^{2}_{1} + {\beta^2\over 4g^2}\, (\varphi^{3}_{0}- \mbox{Re} 
\varphi^{3}_{2}) ,\label{a2}
\end{eqnarray}
Equations (\ref{a1}) imply that $(\mathbb{L}\psi)^m_{0}=0$ for $m=0, 3$ in (\ref{67}).
Since $\varphi^0_{0}= (n^++n^-)/2$ and $\varphi^3_{0}=(n^+-n^-)/2$, these conditions 
yield
\begin{eqnarray}
&& \left. {\tau\over 2}\, {\partial (n^+ + n^-)\over\partial t}\right|_{0}
- {\alpha\tau\over\hbar}\,\Delta^- \mbox{Im}\varphi_{1}^0
- {\gamma\tau\over\hbar}\,\Delta^- \mbox{Im}\varphi_{1}^3= 0,\nonumber\\
&& \left. {\tau\over 2}\, {\partial (n^+ - n^-)\over\partial t}\right|_{0}
+\delta_{2} n^+ - {\alpha\tau\over\hbar}\,\Delta^- \mbox{Im}\varphi_{1}^3
- {\gamma\tau\over\hbar}\,\Delta^- \mbox{Im}\varphi_{1}^0= 0,\nonumber
\nonumber
\end{eqnarray}
wherefrom we obtain
\begin{eqnarray}
A_{0}^\pm = \mp {n^+\over\tau_{\rm sc}} + {\alpha\pm\gamma\over\hbar}\,\Delta^-
\mbox{Im}(\varphi^0_{1}\pm\varphi^3_{1}). \label{a3}
\end{eqnarray}

Let us now calculate $A^\pm_{1}$. Equations (\ref{a2}) imply 
$(\mathbb{L} f^{(2)})^0_{0}=0$ and $(\mathbb{L} f^{(2)})^3_{0}=
f^{(2)\, 3}_{0}$ given by (\ref{a2}) in (\ref{68}). After a little algebra, we find
\begin{eqnarray}
&& A_{1}^\pm = {\alpha\pm\gamma\over\hbar}\,\Delta^- \mbox{Im}(\psi^0_{1}\pm
\psi^3_{1}) -{\beta\over\hbar}\, (\Delta^-\mbox{Re}\psi^2_{1}\pm
\Delta^+\mbox{Im}\psi^1_{1}) \label{a4} \\
&& \quad \mp {\beta\over g\tau}\, \mbox{Im}\psi^2_{1}
\pm {\beta^2\over 8g^2\tau}\, [2\mbox{Re}\varphi^3_{2} +\phi^+_{2}-\phi^-_{2}
-2 (n^+ - n^-)]. \nonumber
\end{eqnarray}

We will now transform (\ref{a4}) in an equivalent form by eliminating Re$\varphi^3_{2}$
and Im$\psi^2_{1}$ in favor of Re$\varphi^3_{2}$ and Im$\psi^2_{1}$, respectively. 
Eq.\ (\ref{66}) implies that $(1+i\vartheta_{2})\varphi^3_{2}= (\phi^+_{2}-
\phi^-_{2})/2$, and therefore,
\begin{equation}
\mbox{Re}\varphi^3_{2}= \vartheta_{2}\,\mbox{Im}\varphi^3_{2}+ {\phi^+_{2}
-\phi^-_{2}\over 2}. \label{a5}
\end{equation}
Similarly, Eq.\ (\ref{67}) implies that $(1+i\vartheta_{1})\,\psi^2_{1}+\delta_{1}\,
\psi^1_{1} =r^2_{1}$, and therefore,
\begin{equation}
\mbox{Im}\psi^2_{1}= -\vartheta_{1}\,\mbox{Re}\psi^2_{1}-\delta_{1}\,
\mbox{Im}\psi^1_{1} + \mbox{Im} r^2_{1}. \label{a6}
\end{equation}
The right hand side of (\ref{67}) yields
\begin{eqnarray}
r^2_{1} = {\beta\over 2g}\,\left({1-e^{-i2kl}\over 2i}\, (\phi^+ -\phi^-)
\right)_{0} - {\beta\tau\over\hbar}\,\Delta^-\left({1+e^{-i2kl}\over 2}\,
\varphi^0\right)_{0}, \nonumber
\end{eqnarray}
wherefrom
\begin{eqnarray}
\mbox{Im}r^2_{1} = {\beta\over 4g}\, (\phi^+_{2} -\phi^-_{2} - n^+ +n^-) - 
{\beta\tau\over 2\hbar}\,\Delta^-\mbox{Im}\varphi^0_{2}. \label{a7}
\end{eqnarray}

Inserting (\ref{a5}),  (\ref{a6}) and  (\ref{a7}) in  (\ref{a4}), we obtain the 
equivalent form:
\begin{eqnarray}
&& A_{1}^\pm = {\alpha\pm\gamma\over\hbar}\,\Delta^- \mbox{Im}(\psi^0_{1}\pm
\psi^3_{1}) -{\beta\over\hbar}\, (\Delta^-\mbox{Re}\psi^2_{1}\pm
\Delta^+\mbox{Im}\psi^1_{1})\label{a8} \\
&& \quad \pm {2\beta\over\hbar}\,\mbox{Im}\psi^1_{1} \pm {\beta\over g\tau}\, 
\vartheta_{1}\mbox{Re}\psi^2_{1}\pm {\beta^2\over 4g^2\tau}\,\vartheta_{2}\,
\mbox{Im}\varphi^3_{2}\pm {\beta^2\over 2\hbar g}\,\Delta^- \mbox{Im}
\varphi^0_{2}. \nonumber
\end{eqnarray}
Inserting (\ref{a3}) and this expression in (\ref{64}) and using (\ref{73}), yield 
(\ref{76}), (\ref{78}) and (\ref{79}). Up to order $\lambda^2$, we have thus proven the 
following statement: 

{\em By using the compatibility conditions in the hierarchy of equations (\ref{67}), 
(\ref{68}), we obtain the same balance equations for $n^\pm$ as by direct substitution of 
the solutions of the hierarchy into equations (\ref{53}) (which arise from integration of the 
kinetic equation over $k$).}

\end{document}